\documentstyle[11pt,newpasp,twoside,epsf]{article}
\begin{document}
\title{The enrichment history of $\omega$~Centauri: what we can learn from 
Str\"omgren photometry}
\author{Michael Hilker}
\affil{Sternwarte der Universit\"at Bonn, Auf dem H\"ugel 71, 53121 Bonn, 
Germany, email: mhilker@astro.uni-bonn.de}
\author{Tom Richtler}
\affil{Departamento de F\'\i sica, Universidad de Concepci\'on, Casilla 160-C,
Concepci\'on, Chile, email: tom@coma.cfm.udec.cl}

\begin{abstract}
In this contribution, results from CCD $vby$ Str\"omgren photometry of a 
statistically complete sample of red giants and stars in the main sequence 
turn-off region in $\omega$~Centauri are presented. From the location 
of stars in the $(b-y),m_1$ diagram metallicities have been determined. 
We argue that the Str\"omgren metallicity in terms of element abundances
has another meaning than in other globular clusters. From a comparison
with spectroscopic element abundances, we find the best correlation with
the sum C+N. The high Str\"omgren metallicities, if interpreted by strong 
CN-bands, result from progressively higher N and perhaps C abundances in 
comparison to iron. We see an enrichment already among the metal-poor 
population, which is difficult to explain by self-enrichment alone.
An attractive speculation (done before) is that $\omega$~Cen was the nucleus 
of a dwarf galaxy. We propose a scenario in which $\omega$~Cen experienced 
mass inflow over a long period of time, until the gas content of its host 
galaxy was so low that star formation in $\omega$~Cen stopped, or 
alternatively the gas was stripped off during its infall in the Milky Way 
potential. This mass inflow could have occurred in a clumpy and discontinuous 
manner, explaining the second peak of metallicities, the abundance pattern, 
and the asymmetrical spatial distribution of the most metal-rich population. 
\end{abstract}

\section{Introduction}

Many medium and high resolution spectroscopy investigations (e.g.
Brown \& Wallerstein 1993; Norris \& Da Costa
1995; Smith et al. 2000) have shown that among the stars in $\omega$~Cen there
exist strong variations of nearly all element abundances investigated so far.
This is reflected by the intrinsic broad scatter of the red giant branch
that cannot be explained by internal reddening only (e.g. Norris \& Bessell
1975). Concerning the iron abundance, several authors have confirmed that 
there exists a main metal-poor population, with a peak at about [Fe/H] $= 
-1.7\pm0.1$ dex, and a broad tail to higher metallicities with a peak at about 
[Fe/H] $= -1.2$ dex (Norris, Freeman, \& Mighell 1996; Suntzeff \& Kraft 1996).
This high metallicity tail extends to values of [Fe/H] $\simeq -0.7$ dex
as deduced from the detection of a very red giant branch (RGB) that is well 
separated from the bulk of the RGB stars (Lee et al. 1999; Pancino et al. 2000).

The abundance variations in $\omega$~Cen point to a more complicated
star formation history than that for other globular clusters (GCs) which
contain a homogeneous stellar population. Whereas the CNO variations might be 
explained by evolutionary mixing effects in the stellar atmosphere as well as 
by mixing in the protocloud (e.g. Bessell \& Norris 1976), the iron abundance 
variations need another explanation (e.g. Vanture, Wallerstein, \& Brown 1994; 
Norris \& Da Costa 1995). An increasing number of groups working on $\omega$~Cen
favour an extended period of star formation connected with self-enrichment as 
the interpretation of their data. Smith et al. (2000, and this volume) studied
the abundances of s-process elements in RGB stars with high resolution 
spectroscopy. A strong increase of [Ba/Fe] and [La/Fe] for metal-poor stars 
with [Fe/H]$<-1.5$ dex followed by a flat relation for higher metallicities 
led the authors conclude that low mass AGB stars have contributed to the 
enrichment. These stars have an evolutionary time of at least 
1 Gyr. Metallicity and age estimates from Str\"omgren $vby$ photometry
in the main sequence turn-off region (Hughes \& Wallerstein 1999, and this 
volume) also suggest an age spread of at least 3 Gyr for stars between $-2.0 
<$ [Fe/H] $<-0.5$ dex (the metal-richest stars also being the youngest). An
age spread also was confirmed in our analysis of Str\"omgren data (Hilker \& 
Richtler 2000), but will not be the main topic in this contribution.
Here, we present the analysis and interpretation of Str\"omgren metallicities 
for more than 1500 RGB stars in $\omega$~Cen. 

\begin{figure}
\plotone{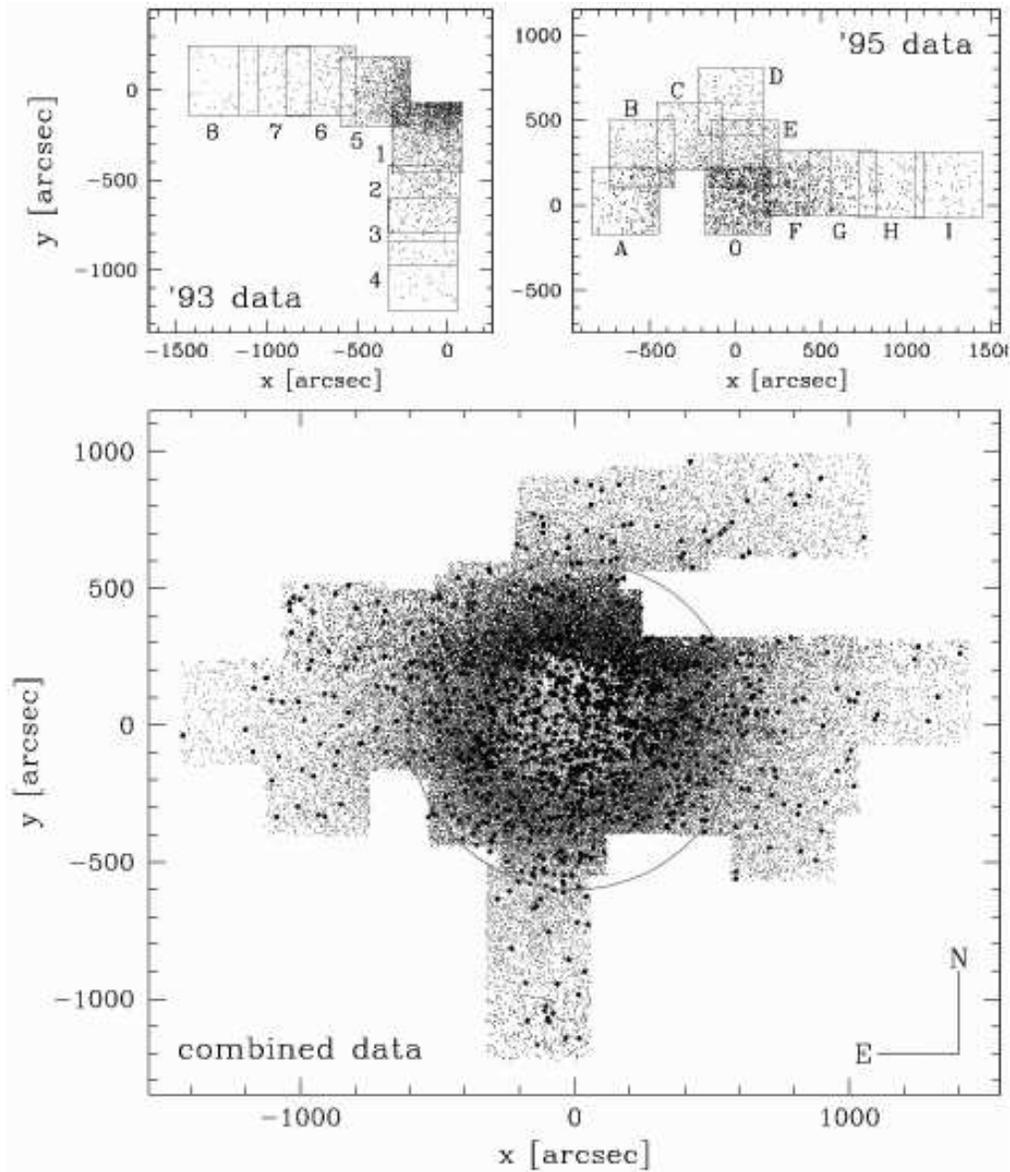}
\caption{The lower panel shows the position plot of all observed
fields in $\omega$~Centauri. All stars with a $V$ magnitude brighter than 19.0
mag and a photometric error less than 0.1 mag have been plotted. Bold dots
indicate the position of all stars with determined metallicities.
The upper panel shows the positions of the long exposures in the '93 run (left)
and the '95 run (right). Only stars with $V < 16.0$ have been plotted.
}
\end{figure}

\begin{figure}
\plotone{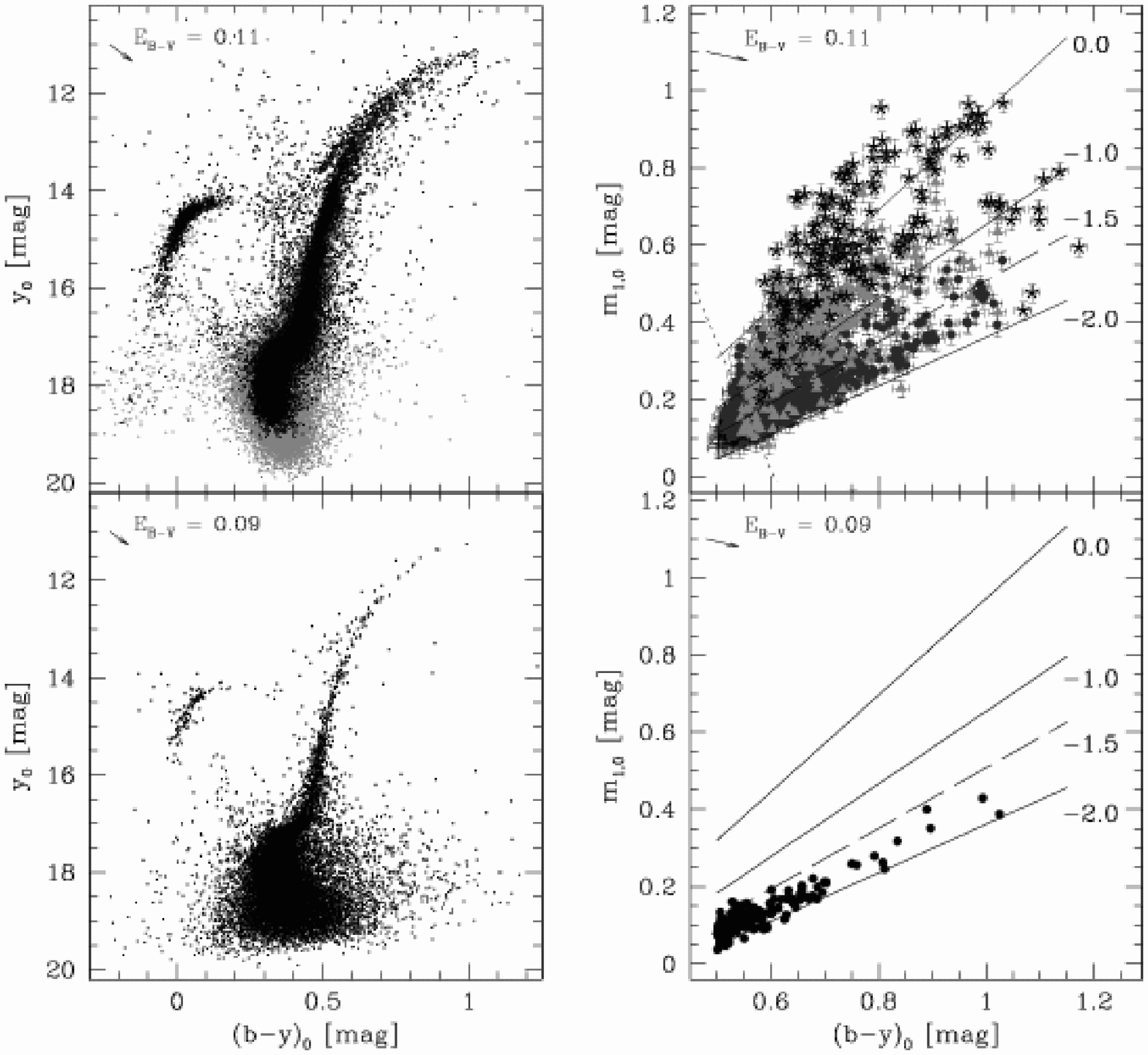}
\caption{The plots on the left show the color magnitude diagrams of
$\omega$~Cen (upper panel) and M55 (lower panel). For $\omega$~Cen, stars
with a photometric error of less than 0.05 mag are marked in grey, the ones
with errors of less than 0.03 mag in black. Whereas the RGB of M55 represent
a single age and metallicity population, the broad red giant branch of
$\omega$~Cen reflects a spread in metallicity and probably age.
In the right panels the $(b-y),m_1$ diagrams for both clusters are shown
together with the lines of constant metallicity from the calibration by Hilker
(2000). The RGB stars of M55 are located on a iso-metallicity line of about
[Fe/H]$=-1.8$ dex (lower panel). In contrast, 1500 selected giants in
$\omega$~Cen show a large spread in their Str\"omgren metallicity due to iron
and CN variations (upper panel). Dark grey dots indicate stars from the blue
side of the RGB, grey triangles the ones from the red RGB side, and asterisks
all stars lying apart from the ``main RGB''. The error bars include photometric
and calibration errors. The dotted line marks the selection criterion
for the metallicity distribution shown in Fig.~3.
}
\end{figure}

\section{Str\"omgren photometry in $\omega$~Centauri}

Str\"omgren photometry has been proven to be a very useful metallicity
indicator for globular cluster giants and subgiants (e.g. Richter, Hilker, \&
Richtler 1999, Hilker 2000, Grebel \& Richtler 1992, Richtler 1989). The 
location of
late type stars in the Str\"omgren $(b-y),m_1$ diagram is correlated with
their metallicities, especially with their iron and CN abundances.
Whereas the color $(b-y)$ is not sensitive to metallicity, the Str\"omgren 
$v$ filter includes several iron absorption lines as well as the CN band at 
4215\AA, and therefore $m_1 = (v-b) - (b-y)$ is a metallicity sensitive index 
(e.g. Bell \& Gustafsson 1978). Within a certain color range, $0.5 < (b-y) 
< 1.1$ mag, the loci of constant iron abundance of giants and supergiants can 
be approximated by straight lines. This is valid for CN-``normal'' ($=$ 
CN-weak) stars. CN-strong stars, due to their higher absorption in the $v$ 
filter, scatter to higher $m_1$ values and therefore mimic a higher 
Str\"omgren metallicity than their actual iron abundance would correspond to.
This can be used to learn more about the CN variations in $\omega$~Cen.
A recent calibration of the Str\"omgren metallicity for CN-``normal'' stars
is presented in Hilker (2000).

The observations of $\omega$~Centauri have been performed in two observing
runs in 1993 and 1995 with the Danish 1.54m telescope at ESO/La Silla.
The details of the observations, data reduction, calibration and photometry
are presented in Hilker (2000) and Hilker \& Richtler (2000). The positions 
of all observed fields are illustrated in Fig.~1.

\subsection{Color magnitude diagram and two-color diagram}

In Fig.~2 the color magnitude diagrams (CMD) and two-color diagrams of 
$\omega$~Cen and M55 are plotted. In the CMD of $\omega$~Cen, all stars with a
photometric error less than 0.05 mag (grey dots) and less than 0.03 mag
($\simeq$ 20620 stars, black dots) in $y$ and $b$ are shown.
The colors in all plots have been corrected for reddening with a value of
$E_{B-V} = 0.11$ mag for $\omega$~Cen (Zinn 1985; Webbink 1985; Reed, Hesser,
\& Shawl 1988; Gonzalez \& Wallerstein 1994), and $E_{B-V} = 0.09$ mag for 
M55 (the mean value between Harris 1996 and Richter et al. 1999).

In both diagrams the difference between a single age and metallicity cluster,
as M55, and the unusual cluster $\omega$~Cen can be seen very nicely.
The broad red giant branch of $\omega$~Cen cannot be explained by photometric
errors or internal reddening. Since the $(b-y)$ color 
is not affected by CN variations, the spread in the RGB
must be due to a spread in the overall metallicity (iron abundance)
and/or age.

The $(b-y),m_1$ diagram (Fig.~2, upper right panel) is indicative for the 
metallicity distribution and CN variations of the red giants in $\omega$~Cen.
The 1500 red giants that have been selected from the CMD show a large scatter
between $-2.0$ and 1.0 dex in their Str\"omgren metallicity.
The Str\"omgren metallicity is defined as
\begin{equation}
{\rm [Fe/H]}_{\rm ph} = \frac{m_{1,0} - 1.277 \cdot (b-y)_0 + 0.331}{0.324
\cdot (b-y)_0 - 0.032}
\end{equation}
following the calibration by Hilker (2000).
The trend exists that stars on the blue side of the RGB are mostly
metal-poor, whereas the stars redwards of the ``main'' RGB populate the
metal-rich regime.

\begin{figure}
\plotone{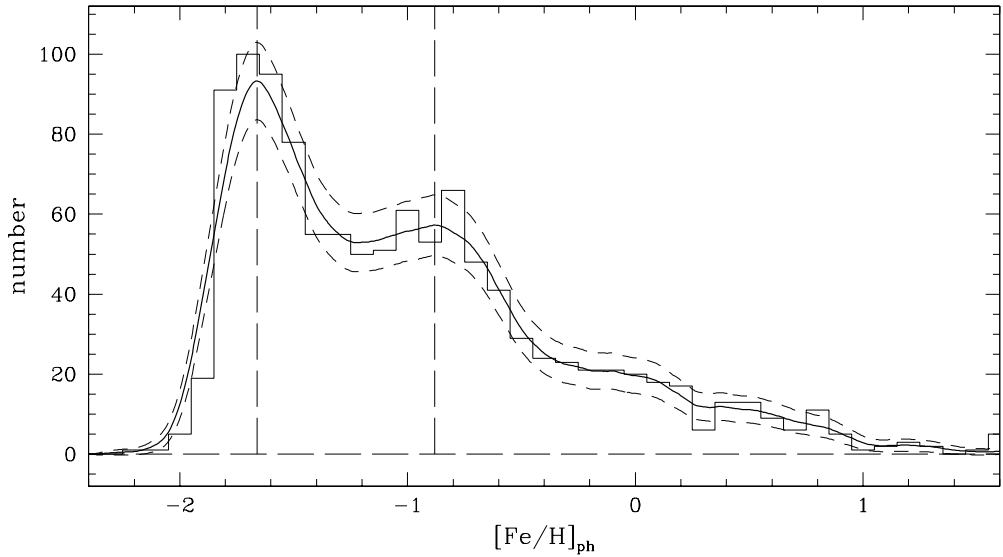}
\caption{
In this plot the metallicity distribution of 1120 selected RGB stars 
(stars redder than the dotted line in Fig.~2) is shown. Stars from the blue 
side of the RGB peak around $-$1.7 dex. Red RGB stars peak around $-0.9$ dex, 
whereas most of the ``reddest'' stars scatter to higher metallicities. 
Note that most of the stars with metallicities higher than $-$0.8 stars are 
CN-rich stars.
}
\end{figure}

\subsection{The metallicity distribution}

In Fig.~3 the metallicity histogram of RGB stars with an accurate metallicity
determination (stars redder than the dotted line in Fig.~2) is presented.
[Fe/H]$_{\rm ph}$ denotes the Str\"omgren metallicity. A peak around $\simeq
-1.7$ dex with a sharp cutoff towards low metallicities at $-1.9$ dex
represents the blue RGB stars. Also most of the AGB stars bluewards the main
RGB belong to this metal-poor population.
Stars from the red side of the RGB have metallicities mainly in the range
$-1.3$ to $-0.5$ dex, with a probable second peak at about $-0.9$ dex.
Our metallicity distribution resembles fairly well the results of Norris et al.
(1996). Stars with Str\"omgren metallicities higher than about
$-0.8$ dex are supposed to be CN-rich stars of one of the two
populations, since no stars with an iron abundance higher than that has
been found in the cluster. Most of them are redder than the ``main'' RGB,
thus belong to a more metal-rich population.
When selecting the RGB stars by a cut in the CMD that corresponds to
a cut in their mass function (more ore less a luminosity cut), the
proportion of metal-poor to metal-rich stars is about 3:1.

\section{Str\"omgren metallicity versus Fe, C and N abundance}

The metallicity distribution found in our investigation is qualitatively
very similar to that found by Norris et al. (1996) and Suntzeff \& Kraft (1996) 
from their Calcium abundance measurements. In Fig.~4 we show the metallicity 
distribution of those stars that are in common in Suntzeff \& Kraft's and our 
samples. Their Calcium abundances have been transformed to iron abundances 
according to the relation given in their paper.

The behaviour of $\omega$~Cen regarding its relation between Fe abundance
and Str\"omgren metallicity is remarkably different from that of other
globular clusters (open circles in the lower panel of Fig.~4),
including NGC~6334, NGC~3680, NGC~6397, Melotte~66, M22 and M55, taken from
Hilker (2000).

\begin{figure}
\plotone{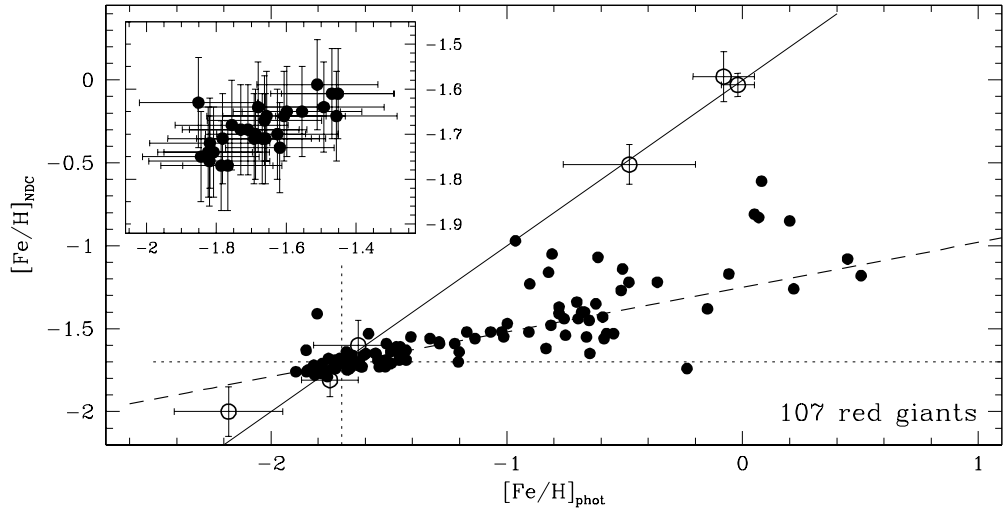}
\caption{
Str\"omgren metallicity versus the iron abundances of those 
red giants that are in common with the sample of Suntzeff \& Kraft (1996). 
The solid line is the iso-metallicity 
line. Whereas most stars of the metal-poor peak and the  
``normal '' GCs (open circles) are located close to 
this line, basically all more metal-rich stars clearly deviate from this 
relation in a systematic manner (dashed relations). 
This relation already is present among the metal-poor stars (upper left).
}
\end{figure}

The straight relation up to -1 dex (with large scatter towards higher
metallicities) is in striking contrast, for example, to the
situation in M22 (Richter et al. 1999), where there is a considerable
scatter at a fixed iron abundance. This relation already is present among
the metal-poor stars (see small plot in the upper left of Fig.~4). No
systematic effect that could cause this relation has been determined. It is
not dependent on a magnitude, color or error selection.
So, what determines the Str\"omgren colors in $\omega$~Cen?
An answer may come from a comparison of the available elements abundances for
40 giants from Norris \& Da Costa (1995).
Fig.~5 shows in four panels (on the left) the
Str\"omgren metallicity vs. [Fe/H]$_{\rm sp}$, [C/H], [N/H], and [C+N/H]. It 
is apparent that the correlation with [Fe/H]$_{\rm sp}$ and [C/H] is very 
poor. It is better for
[N/H] (note the large error of 0.4 dex given for the N-abundance by Norris \&
Da Costa (1995)), and best for [C+N/H].

On the other hand, there is a close correlation of [Fe/H]$_{\rm sp}$ vs. 
[C+N/H]
(Fig.~5, right panel) (which, by the way, is suprising, given the above large
error of the N abundance). The two most deviating stars are ROA 139 and ROA
144. They have the highest N-abundances in this sample, simultaneously low
oxygen abundances and hence are probably strongly affected by mixing effects.
If we skip them, a linear regression returns $0.64 \pm 0.07$ for the slope,
indicating that the increase in C+N is faster than in [Fe/H]. 

What can we learn from the C+N variations in $\omega$~Cen? Can they be
understood as a {\bf stellar evolutionary effect}? 

As Norris \& Da Costa point out, C-depletion as a signature of the CNO-cycle 
is present and one may see the increase in the C-abundance with [Fe/H] in 
their Fig.~8a to have its cause in the decreasing efficiency of the mixing-up 
of processed material with increasing metallicity (and thus mimicking a 
C-enhancenment), as it is theoretically 
expected (e.g. see Kraft 1994). But then, the increase
in C+N is not easy to understand, since it is dominated by an increase of N,
where we would expect a decrease, and, after all, the sum of C and N should be
less sensitive to mixing effects.

\begin{figure}
\plotone{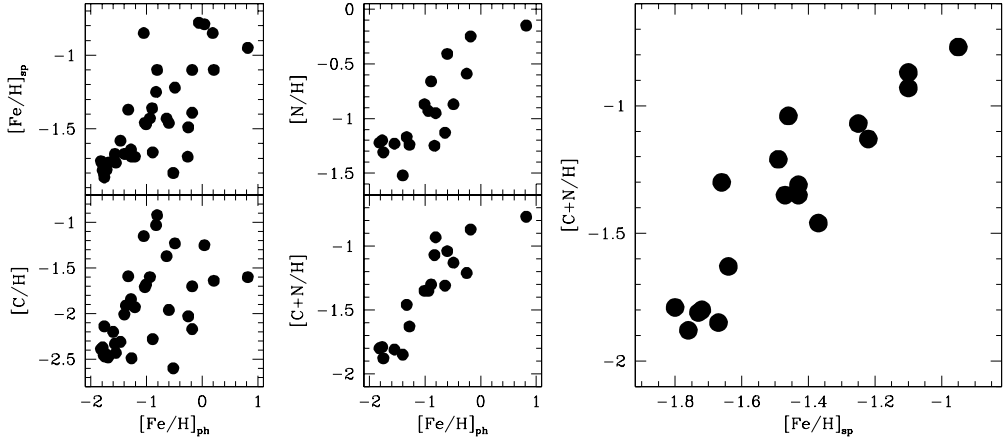}
\caption{In the left,
the spectroscopically determined iron, nitrogen, carbon and nitrogen+carbon
abundances of red giants in $\omega$~Cen (Norris \& Da Costa 1995)
are plotted versus their Str\"omgren metallicity. Whereas the correlation
with [Fe/H] and [C/H] apparently is very poor, it is better for
[N/H] (Norris \& Da Costa give an error of 0.4 dex fot their N-abundances)
and best for [C+N/H]. In the right, [Fe/H]$_{\rm sp}$ is plotted versus the sum of
nitrogen+carbon abundances of red giants in $\omega$~Cen. The slope of
$0.64 \pm 0.07$ indicates a faster increase in [C+N/H] than in [Fe/H]. See
text for further comments.
}
\end{figure}

So, is this a {\bf primordial effect}? (We use the term ``primordial'' for
pre-enriched material as the alternative to mixing effects).

A striking fact for this possibility is that we see in Fig.~4 the relation 
between [Fe/H] and Str\"omgren metallicity {\it already present among the old 
population}, where it is
hard to understand that such small differences in [Fe/H] would cause a regular
pattern in the mixing effects. We thus propose that the gradual enrichment of
C+N, indicated by the Str\"omgren colors, is to a large degree primordial.

The suggestion that a part of the proto-cluster material of $\omega$~Cen has
undergone considerable C-enrichment has also been made by e.g. Cohen \& Bell
(1986) and Norris \& Da Costa (1995) based on the unique
presence of CO-strong stars.
However, the [C/Fe] abundance in the metal-poor population is about $-$0.7 dex
according to Norris \& Da Costa, which is close to $-$0.5 dex, theoretically
expected from the yield ratios in SNe II (Tsujimoto et al. 1995).
Also a mean [O/C]-value of approximately
$-$1 dex, as one would read off from Norris \& Da Costa is close to the
expected yield ratios.

If, say, $-$2.2 dex\footnote{note that 
in our paper (Hilker \& Richtler 2000) erranously the value $-$0.2 dex was
given} would be the ``starting'' value of [C/H], one would require
a considerable C-contribution from intermediate-age stars, which by itself is 
not easy to understand for the first stars formed in $\omega$~Cen. 
Therefore, if the increasing [C+N/Fe] among the metal-poor old population was, 
at least to large part, primordial, one is driven to the conclusion that 
already the star formation process, which formed these stars, did not took 
place in a single burst within a well-mixed environment, but must have been 
extended in time, allowing intermediate-age populations to contribute.

We shall attempt to combine these abundance pattern with other properties
within a scenario later on.

\section{Str\"omgren metallicity versus other properties in $\omega$~Centauri}

As already mentioned in the introduction the metallicity spread is 
accompagnied by an age spread. Metallicity determinations and isochrone
fitting of stars in the main sequence turn-off region (Hughes \& Wallerstein 
1999; Hilker \& Richtler 2000) revealed 
that the more metal-rich stars tend to be younger. Whereas all stars of the 
main RGB with metallicities between $-2.0$ and $-1.4$ dex might be compatible 
with one age, the populations with metallicities around $-1.2$ and $-0.7$ dex 
are at least 2--4 and 3--5 Gyr younger. If the age metallicity relation in 
$\omega$~Cen can be understood as a continuous enrichment process after an 
initial starburst with [Fe/H]$\simeq-1.7$ dex, the age spread of the enrichment 
lies between 3 and maximally 5 Gyr. 

\begin{figure}
\plotone{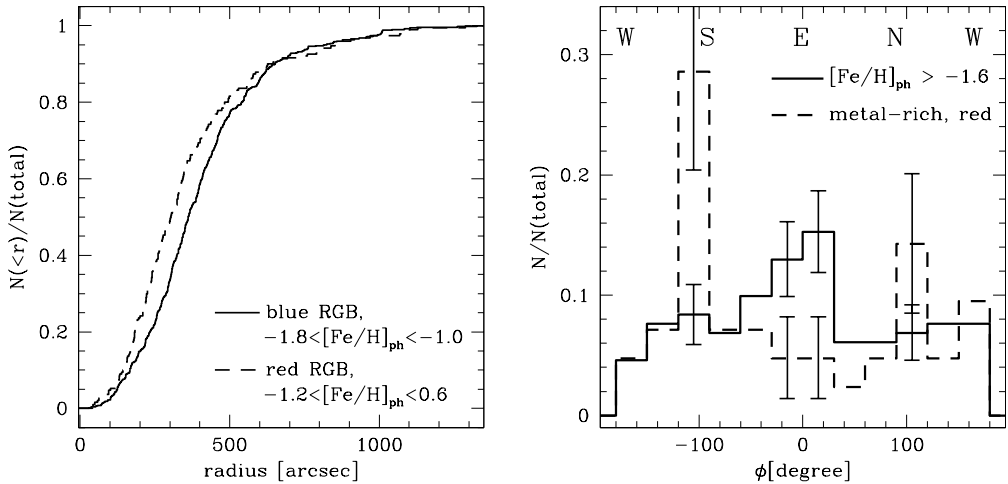}
\caption{Left: cumulative distribution of different 
subsamples of red giants in a 20$\arcmin$ broad East-West strip. The 
metal-richer stars are more concentrated than the 
metal-poor population, with a significance of 99.9\% (KS test).
Right: angular distributions of the most metal-poor 
and most metal-rich giants in $\omega$~Cen.
The number counts are normalized to the total number in each sample.
Error bars indicate the statistical error in the corresponding bins.
An excess of very metal-rich stars in the South direction is evident.
}
\end{figure}

\subsection{Spatial distribution of sub-populations}

To investigate the spatial distribution of sub-populations in $\omega$~Cen,
the 1500 red giants with metallicity determinations have been divided
into four sub-populations according to their age and metallicities:
(1) an old metal-poor population from the blue side of the RGB, (2) a more
metal-rich, mostly CN-rich, and younger population from the red side of 
the main RGB, (3) the youngest, very metal- and CN-rich population, and (4)
the AGB stars of the old population. 

Both, the cumulative radial distributions as well as the angular distributions
of the sub-populations have been examined. For the angular distribution,
only stars within a radius of $10\arcmin$ from the cluster center have been
included. The number counts have been normalised to the total number for each 
selection. The angle $\phi$ is defined as $0\deg$ in East direction, 
$+90\deg$ North, and $-90\deg$ South.
 
The population of the selected 83 AGB stars, which is very metal-poor, is 
distributed as the metal-poor pop (1), as expected. 
Deviations in the angular distribution are statistically not significant.

When comparing population (1) with population (2) a difference 
in their radial distributions becomes evident in the sense that the ratio
of the metal-rich to metal-poor stars is higher in the cluster center
than in its outskirts (see Fig.~6, left panel). This result is statistically 
significant. A KS test reveals a probablity of less than 0.1\% that the 
cumulative number counts of both populations follow the same radial 
distribution.  

The radial distribution of the stars in pop (3) appears less
concentrated than the average cluster population. Some of them might
be solar metallicity foreground stars. However, others are confirmed cluster
member stars. The angular distribution of pop (3) shows a concentration of
stars towards the South and a slight depression in the West and North
direction (see Fig.~6, right panel) which can explain the different radial 
distribution. We note that also Jurcsik (1998) reported on a spatial 
metallicity asymmetry in $\omega$~Cen. She found that the most metal-rich 
stars with [Fe/H]$>-1.25$ dex are concentrated towards the South, whereas the
most metal-poor stars with [Fe/H]$<-1.75$ are more concentrated in the North.
As shown in Fig.~6, nearly 30\% of our metal-rich sample are located in the 
Southern angular bin, but an asymmetrical distribution in the North-South
direction of the most metal-poor stars cannot be confirmed. The probability 
that the angular distribution of the metal-poor stars of Jurcsik's and our 
sample agree is less than 7\% (KS test). Also less than 7\% is the probability 
that metal-rich and metal-poor stars are distributed equally.

\subsection{Metallicity and kinematics}

Along with the abundance variations in $\omega$~Cen, different sub-populations
show clearly different kinematical behaviour. A dynamical analysis of 400 
stars in the Norris et al. (1996) sample of RGB stars with calcium 
abundance measurements revealed a rotation of the metal poor component, 
whereas the metal rich one is not rotating (Norris et al. 1997) 
Recent studies with larger radial velocity samples, as presented in this
conference proceedings (Gebhardt et al.; 
Seitzer et al.; Cannon et al.) confirm this result.

A match of our data set with that of Xie, Pryor \& Gebhardt (private
communication) shows similar results. Within 3$\arcmin$ radius we have more 
than 1000 RGB stars in common. Their radial velocity dispersion peaks in the 
center at about 19 km/sec, falling to 16 km/sec at a radius of 2$\arcmin$.
When dividing the sample in 860 metal-poor ([Fe/H]$_{\rm ph}<-0.6$) and 150
metal-rich stars, one clearly can see differences in their rotation behaviour.
Whereas metal-poor stars show a strong sign of rotation around the 
semi-minor axis of the elliptical shape of $\omega$~Cen, no rotation is seen
among the most metal-rich stars. At 3$\arcmin$ radius from the center the
rotation velocity of the metal-poor stars reaches 3.6 km/sec.
Further detailed anlysis of both data sets will show what more can be learned
about the connection between chemical enrichment and kinematical behaviour.

\section{A scenario}

\subsection{Problems with self-enrichment within $\omega$~Centauri}

Can the younger populations in $\omega$~Cen be enriched by the older one?
In trying to demonstrate the problems with this picture we use oxygen as a 
tracer for the synthesized material. First we estimate the number of SNe 
type II having occured in the old population. If we adopt the mass of 
$\omega$~Cen to be $4\times10^6$ solar masses (Pryor \& Meylan 1993), the 
metal-poor population comprises about $2.8\times10^6$ M$_{\sun}$. 
We get 45000 SNe for all stars more massive than 10 M$_{\sun}$ when assuming 
a Salpeter mass function between 0.1 and 100 M$_{\sun}$.
The total oxygen mass released by these SNe is about 50000 M$_{\sun}$
(based on Table 7.2 of Pagel 1997). On the other hand, following Norris \& Da 
Costa (1995), a mean [O/H] value for the metal-rich population is $-$0.7 dex 
(adopting [O/Fe] = 0.5 dex and [Fe/H] = $-$1.2 dex), so we calculate the actual 
oxygen mass to be 2300 M$_{\sun}$, if the total mass of the young population 
is $1.2\times10^6$ M$_{\sun}$. Since Smith et al. (2000) see no signature of 
enrichment by SNe Ia up to [Fe/H] = $-$0.9 dex, it is reasonable to assume 
that the oxygen mass, which was present
already in the gas before the enrichment, scales with the iron abundance.
The ratio is about a factor of 3, so we have 1500 M$_{\sun}$
of newly synthesized oxygen in the metal-rich population. This
means that only 3\% (or less) of the released oxygen has been
retained. If we do this exercise with the iron abundance, we have 1000 
M$_{\sun}$ of iron released (Pagel, p.~158), and we have about 20 M$_{\sun}$
of newly synthesised iron present. Of course, the exact numbers
are insignificant, but the above consideration suggests that practically 
{\it all} material must have been blown out.

This is also plausible from the energy point of view. We have a release of
kinetic energy of about $5\times10^{54}$ erg from the SNe (neglecting previous
stellar winds and ionizing radiation), while the binding energy of the
``proto-young population'' is about $2\times10^{52}$ erg, if for simplicity we
imagine that the gas was confined within a half-light radius of 7 pc
(Djorgovski 1993).

Similar factors must apply to the overall fraction of retained gas, implying
an unreasonably large protocluster mass (neglecting the problem of how a bound
system could survive after such strong mass loss), if one wants to keep the 
hypothesis of a permanently retained large gas fraction within $\omega$~Cen.

But we have even more problems. Smith et al. (2000) only detect
weak signatures of SNe Ia in the metal-rich population, expressed by the low 
[Cu/Fe] value of $-0.6$ dex.
On the other hand, the age spread, the increase of s-process elements,
and the interpretation of the
Str\"omgren results as primordial enrichment of C and N speak for the
contribution of an intermediate-age population. Within 2-3 Gyr, at least some
Ia events should have ocurred, making the problem with the low iron
content even worse, if they would have provided iron to the young population.
Why do we not see their debris?

Moreover, it is remarkable that we find the signature of intermediate-age
populations already among the old population, indicated by the evidence that
the same relation between [Fe/H] and Str\"omgren metallicity (Fig.~4), which
connects the oldest and the younger population, appears already to be present
at the lowest metallicities. In this respect, the general behaviour of the 
Str\"omgren metallicities resembles the well established enrichment of
s-process elements relative to iron. 
How can these stars, in a regular pattern, be self-enriched simultaneously by 
SNe II and by intermediate-age stars?

It may be that one can construct a scenario in which these oddities can be
explained by pure self-enrichment (i.e. Smith et al. 2000). However, we wish 
to point out an alternative, which seems to offer an easier way towards an
understanding of $\omega$~Cen.

\subsection{Just imagine ...}

... $\omega$~Cen formed within a formerly much larger entity, outside the
Milky Way, and at the central position of its ex-host-object.
We (speculatively) assume that this object was a dwarf galaxy with
$\omega$~Cen as its nucleus.
We additionally speculate that its star formation rate was triggered
over a very extended period (perhaps more than 5 Gyr) by {\it mass supply}
from the overall gas reservoir of its
host galaxy. This scenario can explain all characteristic properties of
$\omega$~Cen found so far.

This gas inflow, already enriched in the host galaxy to at least $-0.9$ dex, 
could have occured in
a non-spherical, clumpy and discontinuous manner, providing angular
momentum to the first population in $\omega$~Cen and thus giving rise to the 
flattening of $\omega$~Cen. That no significant rotation is seen in the 
more metal-rich population might be due to the loss and transfer of angular 
momentum from the newly infalling gas to the very massive rotating dark matter 
halo of the first population (see ideas by Binney, Gerhard, \& Silk 2001). 
We have no
problems with the competition of gas removal and simultaneous enrichment.
The intermediate-age population stars in $\omega$~Cen released their gas,
for instance by planetary nebulae, in a much less violent fashion and
the infalling gas mixed with this C and N rich material, which also
was rich in s-process elements, giving rise to a new star formation period.
The large scatter in the Str\"omgren metallicities may thus be in part
primordial, reflecting the incomplete mixing of the infalling gas with the
C-N-rich material.
Both SNe II,Ib and Ia would sweep up the gas almost completely, terminating
star formation for a short while, until further mass infall becomes possible.

It also seems natural that younger and more metal-rich populations show other
kinematic and spatial properties, including asymmetries in their spatial
distribution, depending on the details of the infall process. The initially 
asymmetrical distribution of the most metal-rich population is probably still
not relaxed due to the long relaxation time of $\omega$~Cen beyond the 
half-mass radius (Meylan 1987).
We would then expect many periods of strong star formation alternating with
periods of mass infall. The mass infall would finally cease after the gas
content of the host galaxy has become sufficiently low or was perhaps removed
by ram pressure stripping in the Galactic halo during its infall in the
Milky Way (e.g. Blitz \& Robishaw 2000). 

The subsequent evolution can be sketched as follows: on its retrograde
orbit the dwarf galaxy spiralled towards the Galactic center (Dinescu, Girard,
\& van Altena 1999). On its way it lost the outermost stellar populations
by tidal stripping, including the likely member globular clusters
NGC 6779 (Dinescu et al. 1999). Finally, after
its stellar population dissolved totally, the nucleus $\omega$~Cen
remained and appears now as the most massive cluster of our Milky Way.

\section{Summmary and concluding remarks}

For about 1500 red giants in $\omega$~Centauri, Str\"omgren metallicities
have been determined. Almost 2/3 of them turn out to be metal-poor, with a peak
at [Fe/H]$_{\rm ph} = -$1.7 dex. Beyond this peak, the metallicity distribution
shows a sharp cutoff towards lower metallicities, but a broad, long tail
towards higher metallicities. Most of these stars are CN-rich and
have ages 2-5 Gyr younger than that of the oldest population.

The comparison between [Fe/H] abundances derived from high-dispersion
spectroscopy of Norris \& Da Costa (1995) and
Str\"omgren metallicities shows a behaviour distinctly different from that
observed in other globular clusters. There is hardly a correlation with
[Fe/H], but a close correlation with [C+N/H].

However, the comparison of Str\"omgren metallicities to the larger sample
of Suntzeff \& Kraft (1996) shows that there is a coupling to
the iron abundance indicating that this has a primordial cause. It is
already visible among the metal-poor population and we interpret it as
another manifestation of the well established increasing contribution
of intermediate-age populations with increasing iron abundance.

The comparison of the cumulative radial distribution of the two main
populations in $\omega$~Cen exhibits a higher concentration of the metal-rich
stars. The youngest, most metal-rich population has an asymmetrical 
distribution around the cluster center with a concentration towards the South.

When combining the kinematics of more than 1000 red giants with our Str\"omgren 
metallicities, a significant rotation of the metal-poor population was 
confirmed, whereas the most metal-rich stars do not rotate.

Our findings are consistent with a scenario in which enrichment
of the cluster has taken place over a period of 3--6 Gyr.
The conditions for such an enrichment can perhaps be
found in nuclei of dwarf galaxies. All characteristic
properties of $\omega$~Cen (flattening, abundance pattern, age spread,
kinematic and spatial differences between metal-poor and metal-rich stars)
could be understood in the framework of a scenario, where infall of
previously enriched gas occured in $\omega$~Cen over a long period of time.
Only the enrichment of nitrogen, carbon, and s-process elements took place
within $\omega$~Cen, where the infalling gas mixed with the expelled matter
from AGB stars.

The capture and dissolution of a
nucleated dwarf galaxy by our Milky Way and the survival of $\omega$~Cen
as its nucleus would thus be an attractive explanation for this extraordinary
object. Several contributions in this conference proceedings also support this 
idea and rule out other possibilities like a chemically diverse parent cloud 
or a merger of two clusters.

\acknowledgements We thank Bingrong Xie for providing us radial velocities
from their Fabry-Perot sample (see also Gebhardt, this volume). This project 
was partly supported through `Proyecto FONDECYT 3980032'.

\end{document}